\documentclass{aa}
\usepackage{graphicx,epsfig}
\def\etal{{\it et al}}

\def\P3hat{{\mathaccent 94 P}_3}

\def\eg{{\it e.g.}}
\def\ie{{\it i.e.}}

\newcount\notenumber
\def\clearnotenumber{\notenumber=0}
\def\note{\advance\notenumber by1 \footnote{$^{\the\notenumber}$}}
\clearnotenumber
\begin{document}

\title{The Topology and Polarisation of Subbeams Associated with the `Drifting' Subpulse Emission of Pulsar B0943+10 --- IV. Q-to-B-Mode Recovery Dynamics}

\author{Joanna M. Rankin\inst{1}\thanks{On leave from Physics Dept., 
University of Vermont, Burlington, VT 05405 USA, email: joanna.rankin@uvm.edu} 
   \and Svetlana A. Suleymanova\inst{2,3}
\offprints{joanna.rankin@uvm.edu}
\institute{Sterrenkundig Instituut
   `Anton Pannekoek', 1098 SJ Amsterdam, The Netherlands
   \email{jrankin@astro.uva.nl}  \and Pushchino Radio Astronomy
   Observatory, 142292 Pushchino, Russia \email{suleym@prao.psn.ru}
   \and Isaac Newton Institute of Chile, Pushchino Branch} }
%\begin{document}

\abstract{Pulsar B0943+10 is well known for its `B' (burst) mode, characterized by accurately drifting subpulses, in contrast to its chaotic `Q' (quiet) mode. Six new Arecibo observations at 327 MHz with  durations of 2+ hours each have shed considerable light on the modal dynamics of this pulsar. Of these, three were found to be exclusively `B' mode, and three were discovered to exhibit transitions from the `Q' to the `B' mode. One of these observations has permitted us to determine the circulation time of the subbeam carousel in the `Q' mode for the first time, at some 36.4$\pm$0.9 stellar rotation periods. The onset of the `B' mode is then observed to commence similarly in all three observations. The initial circulation time is about 36 periods and relaxes to nearly 38 periods in a roughly exponential fashion with a characteristic time of some 1.2 hours. This is the longest characteristic time ever found in a mode-switching pulsar. Moreover, just after the `B'-mode onset the pulsar exhibits a symmetrical resolved-double profile form with a somewhat stronger trailing component, but this second component slowly dies away leaving the usual single `B'-mode profile with the longitude of the magnetic axis falling at about its trailing half power point.  Thus it would appear that Q-to-B- and B-to-Q-transitions have different characteristic times. Some speculations are given on the nature of this slow modal alternation. 
}

\date{Received / Accepted }
\authorrunning{Rankin \& Suleymanova}
\titlerunning{Pulsar 0943+10's `Drifting'-Subpulse Emission IV.}
\maketitle

\section*{I. Introduction}

Pulsars are known to form very stable intensity profiles after averaging hundreds of individual pulses. Additionally, there are a number of peculiar, so-called ``mode switching'' pulsars, which have two stable integrated-profile shapes. One prominent such pulsar is B0943+10. Several studies have described the emission characteristics of its bright `B' mode and less intensive (in average) `Q' mode (Suleymanova \& Izvekova 1984; Suleymanova \etal\ 1998, hereafter SIRR).  

In 1992 a nearly 1000-pulse sequence was recorded using the Arecibo radio telescope at 430 MHz, and it was found to begin with some 800+ pulses in the B mode followed by nearly 200 pulses in its Q mode. The transformation process from the B to Q mode was thoroughly investigated. It was shown that the Q-mode onset resulted from both abrupt and gradual changes in the intensity of individual pulses. The presence of slow changes in the modal transformation process having a timescale of some 20 minutes were documented for the first time. 

This key Arecibo observation was also used to determine the subbeam configuration responsible for the drifting-subpulse pattern for the first time (Deshpande \& Rankin 1999; 2001, hereafter Paper I), and 35-MHz Gauribidanur observations revealed the same B-mode pattern at very low frequency (Asgekar \& Deshpande 2001, hereafter Paper II).  Furthermore, a remarkable set of simultaneous 40- and 103-MHz recordings made at the Pushchino Radio Astronomy Observatory suggested interesting correlations between the circulation time, profile form and emission mode (Rankin, Suleymanova \& Deshpande 2003, hereafter Paper III), but the short duration of these pulse sequences (hereafter PSs) from transit instruments left many questions unanswered about B0943+10's overall modal dynamics.  

\begin{figure}
\begin{center}
\epsfig{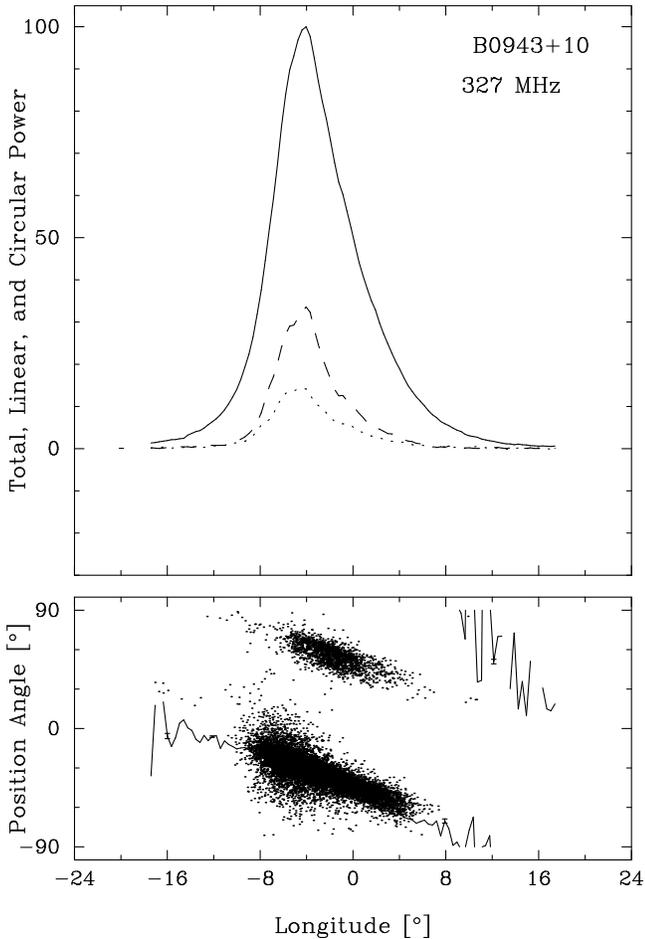}
\end{center}
%\parhang {\bf Figure 1.}---
\caption{Average polarization profile giving the total intensity (solid line), total linear (dashed line) and circular polarization (dotted line) for the B-mode of PSR B0943+10 (top panel).  Linear polarization angles for the integrated profile (solid line) and sufficiently strong individual pulse samples (dots) are given as a function of pulse longitude (bottom panel).  The 327-MHz PS was  recorded on 2003 October 5. 
\label{fig1}}
\end{figure}

\begin{table}
\begin{center}
 \caption{Arecibo 2003 observations at 327 MHz.}
 \begin{tabular}{cccccc}
 \hline
Day & MJD & 2003 & Mode & Switch & Length\\
        &           &           &             &     (pulse)  &  (pulses)\\
 \hline
 1 & 52709 & 10 March & Q to B & 2540 & 6748 \\
 2 & 52711 & 12 March & B         &   ---    & 6809 \\
 3 & 52832 & 12 July     & Q to B & 5266 & 7559 \\
 4 & 52840 &  20 July    & B         &   ---    & 7274 \\
 5 & 52916 & 4 October & Q to B &1755 & 5650 \\
 6 & 52917 & 5 October & B         &   ---   & 3024 \\
\hline
\end{tabular}
\end{center}
\end{table}

It is thus of great interest to further investigate the star's modal transformation processes.  Observations of B0943+10 were carried out using the upgraded Arecibo radio telescope in 2003 at 327 MHz with this aim. As a result, several individual pulse observations have been recorded with durations up to 135 minutes. Remarkably, three of the six observations contain PSs of both modes, all transitions from the Q to the B mode.  One of our main questions was whether the transition process between the star's two modes is symmetrical.  Having heretofore had access to only two short records with modal transitions (B-Q at 430 MHz in Paper I) and (Q-B at 103/40 MHz in Paper III), we now have an excellent opportunity to answer to this question. \S II discusses our observations, \S III compares the properties of the two modes, and \S IV discusses the profile variations incurred in the mode change.  Fluctuation-spectral variations associated with the modal transition are treated in \S V, and in \S VI we attempt to understand where our other B-mode sequences would fall in relation to the modal onset.   \S VII discusses the asymmetry between the B-to-Q and Q-to-B transitions processes, and \S\S VIII and IX give brief discussions and 
summaries of the results, respectively.  
 
\begin{figure}
\begin{center}
\epsfig{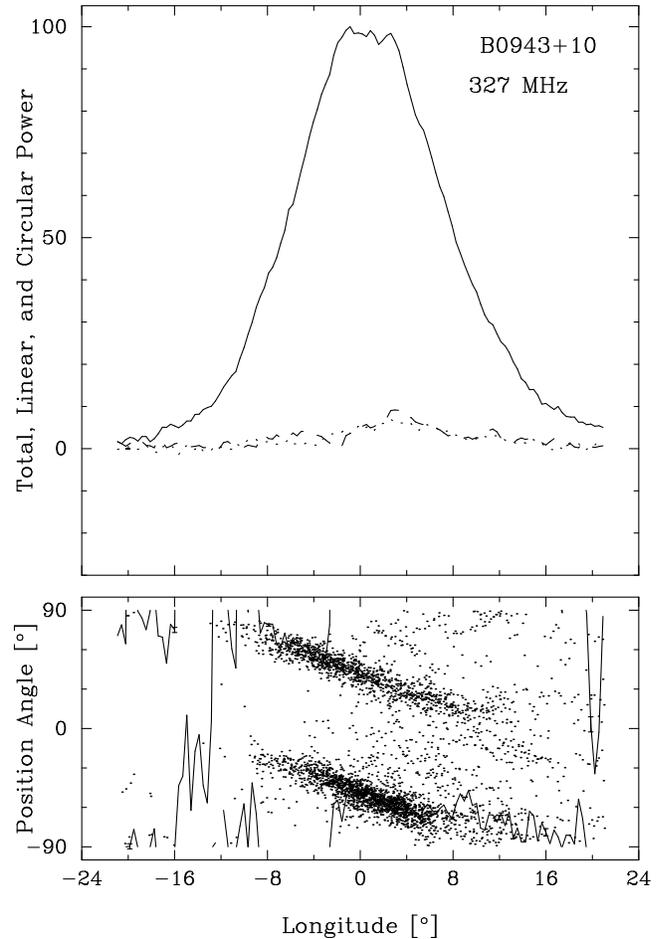}
\end{center}
%\parhang {\bf Figure 2.}---
\caption{Average polarization profile and polarization-angle distribution as in Fig.~\ref{fig1} for the Q-mode of PSR B0943+10. This 327-MHz observation was recorded on 2003 October 4. 
\label{fig2}}
\end{figure}

\section*{II. Observations}

The observations used in our analyses were made using the 305-m 
Arecibo Telescope in Puerto Rico together with its Gregorian reflector 
system and new 327-MHz polarimetric receiver system.  The signals 
were acquired using the Wideband Arecibo Pulsar Processor 
(WAPP\footnote{http://www.naic.edu/$\sim$wapp}) during 2003 as 
detailed in Table 1.  The auto- and cross-correlation functions of the 
channel voltages produced by receivers connected to orthogonal 
linearly polarized feeds were 3-level or 9-level (days 52916/7) sampled.  
Upon Fourier transforming, 64 or more channels were synthesized 
across a 25-MHz bandpass with a 512-$\mu$s sampling time, providing 
a resolution of less than a milliperiod.  The Stokes parameters have 
been corrected for dispersion, interstellar Faraday rotation, and various 
instrumental polarization effects.\footnote{Unfortunately, an error in the 
WAPP software resulted in only one linear polarization being recorded, 
four-fold redundant, during the initial 2003 March observations.}

\section*{III. Characteristics of B0943+10's B  and Q modes}

\begin{figure}
\begin{center}
\epsfig{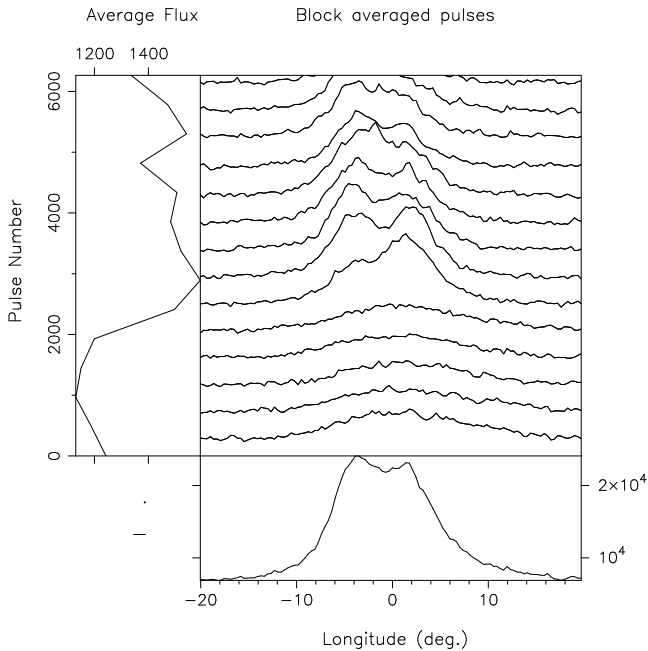}
\end{center}
%\parhang {\bf Figure 3.}---
\caption{A set of total intensity pulse profiles of B0943+10 at 327 MHz (2003 March 10), obtained by integration of 482 consecutive pulses. The first 5 integrations show the smooth, broad and weak profiles of the Q-mode. Note the unusual profile shape just after the burst-like B-mode onset with the strongest emission coming after the magnetic-axis (zero) longitude. The subsequent profiles show the gradually decreasing intensity of the second component . The broad shape of the total integration in the bottom panel shows that Q-like features persist well after the B-mode onset. The intensity integrated over the pulse window as a function of pulse number is shown in the left-hand panel.
\label{fig3}}
\end{figure}

\begin{figure}
\begin{center}
\epsfig{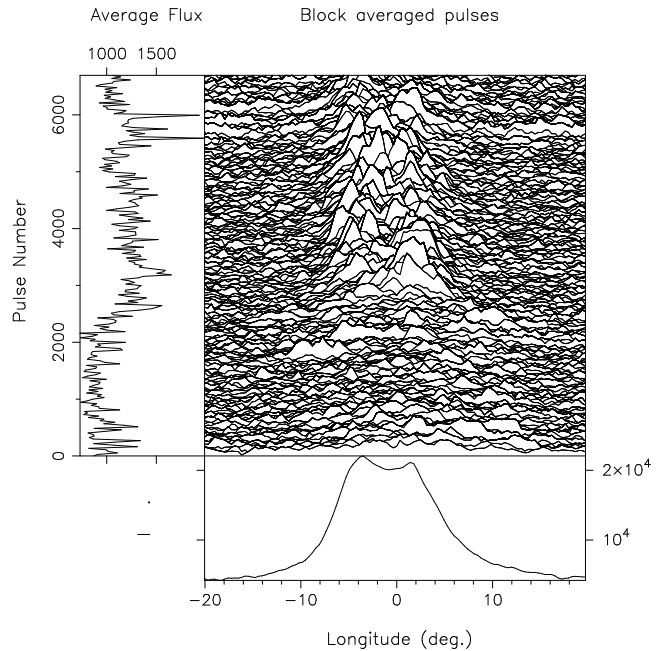}
\end{center}
%\parhang {\bf Figure 4.}---
\caption{Another set of profiles of the B0943+10 sequence in Fig.~\ref{fig3}, obtained here by integration of 27 pulses. The abrupt onset of the B mode at about pulse 2540 is associated with a 50\% increase in the emission intensity and the resumption of subpulse drifting behavior.
\label{fig4}}
\end{figure}

\begin{table}
\begin{center}
 \caption{Modal profile properties of B0943+10 at 327 MHz.}
 \begin{tabular}{cccccc}
 \hline
Mode & Flux & W   &  W  & L/I & V/I \\
           & (arb.)&(ms)&(\degr)&(\%)&(\%)\\
 \hline
 B & 2? & 20 &  7  & 30 & 15 \\
 Q &  1  & 41 & 14 & 10 & 10 \\
\hline
\end{tabular}
\end{center}
\end{table}

Polarized 327-MHz integrated profiles of B0943+10's B and Q modes 
are given in Figures \ref{fig1} \& \ref{fig2}.  These observations were
carried out on 2003 October 5 and 4, respectively. The top panels 
give the total intensity (solid), linear (dashed) and circular (dotted curve) 
polarization, and the lower panels position-angle (hereafter PA) histograms. 
Table 2 summarizes some of the profile properties.

The peak intensity of the Q mode is typically 1/3 that of its B-mode; however 
the Q-mode profile is roughly twice as wide as its B-mode counterpart.  
The second component of the ``pure B''-mode profile is unresolved at 
327 MHz---similar to the situation at 430 MHz (\eg, Paper I) but contrasting 
with that at 103 MHz, where the second component is 1/3 of the amplitude 
of main component and well resolved, its peak trailing at a distance of 
10\degr\ (SIRR).  

The lower panels in Figs.~\ref{fig1} \& \ref{fig2} show the primary- and 
secondary-polarization modes (hereafter PPM and SPM), which form two 
bands of PA sweep in longitude, separated by 90\degr. The B mode 
is dominated by the PPM.  The PA distribution in Fig.~\ref{fig2} is typical 
for the Q mode: the PPM (lower band) and SPM (upper band) span the 
entire longitude range under the pulse window and the sample PAs are 
almost equally distributed between the two modal bands, resulting in a  
nearly complete depolarization of the integrated profile.

\section*{IV. The Q- to B-mode Transitions of B0943+10 at 327 MHz}

\begin{figure}
\begin{center}
\epsfig{file=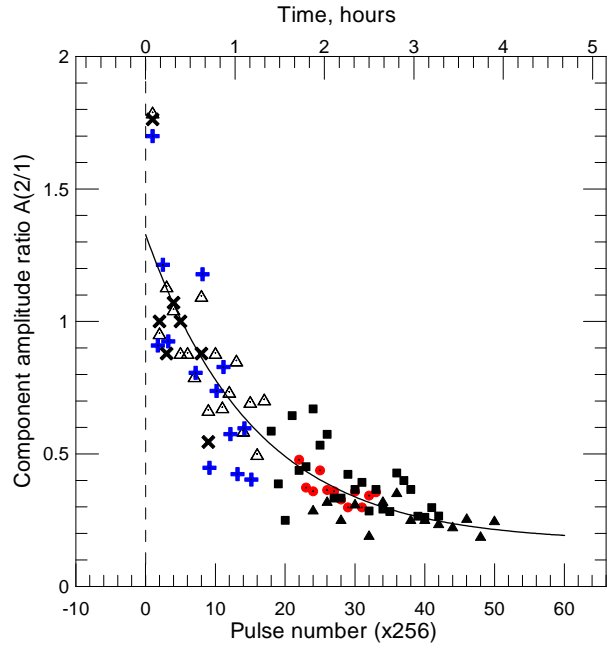,height=8.5cm,angle=0}
\end{center}
%\parhang {\bf Figure 5.}---
\caption{Systematic changes in B0943+10Õs average-profile form as a 
function of time after the B-mode onset. Initially the trailing component 
dominates, but its profile slowly evolves to its usual asymptotic form, where 
the longitude of the magnetic axis falls near the trailing half-power point 
and the intensity of the second component decreases to 1/5 that the first, 
dominant component. Here, we plot the amplitude ratio A(2/1) of the two 
component peaks, separated by 6\degr\ longitude, as a function of time in 
terms of the profile number. Each average is comprised of 256 individual 
pulses, or 280 seconds (4.7 minutes). Note the remarkable similarity of 
the very first B-mode pulse shape with A(2/1) near 1.75 for all three Q- 
to B-mode transition days (hereafter MJD 52709 open triangles, 52832 Xs, 
52916 crosses).  Values for the ``pure B'' days (MJD 52711 squares, 
52840 triangles and 52917 circles) are shifted to begin at profile numbers 
18, 24 and 22, respectively. They clearly show a gradual change of pulse 
shape and seem to represent a later portion of the Q- to B transformation 
process (see also Fig.~\ref{fig10}). This interpretation is further supported 
by the gradually increasing circulation time of all three ``pure B'' days (see 
text and Fig.~\ref{fig8}). The vertical dotted line indicates the B-mode onset 
time.  A fit to the values (solid curve) has the form 0.17$+$1.16 $\exp(-t/\tau)$, 
where $\tau$ is 15.65 256-$P_1$ blocks.  
\label{fig5}}
\end{figure}
%Comment:
%	day		Sign
%	Q/B days	
%	52709 (Q/B1)	Triangles
%	52916 (Q/B2)	Crosses
%	52832 (Q/B3)	dark circles
	
%	Pure B-days	
%	52711 (B1)	Rectangles
%	52840 (B2)	Crosses
%	52917 (B3)	open circles
%For each of Q/B days the first group of 256 pulses is corresponding to the group for which a feature was detected for the first time in fluctuation spectra. All Q/B days show very similar behavior in decreasing of the second component. Ratio of the components asymptotically approach the value 0.25, which is a characteristics of the ``pure'' B-mode well presented by the profile recorded on MJD52840.

We first discovered the Q- to B-mode transition of 2003 March 10, and 
the intensity and profile changes associated with this event are shown 
in Figures~\ref{fig3} \& \ref{fig4}.  The central panel of Fig.~\ref{fig3} gives 
a set of 482-pulse averages, five before the transition and nine afterward.  
The transition occurred at about pulse 2540 as we will discuss further 
below.  Note the initial weak, broad, unimodal Q-mode profile which then 
gives way to a {\em double} B-mode form having a more intense trailing 
component.  This unusual trailing component then weakens progressively 
over the next hour or so, finally resulting in the usual B-mode profile seen 
in Fig.~\ref{fig1}---that is, with an asymmetrical single form that has 
longitude zero (that of the magnetic axis) near its trailing half-power point.  
The left-hand panel gives the integrated intensity of each average, and it 
is clear that the B-mode onset entails an about 50\% increase in the star's 
intensity.  Fig.~\ref{fig4} gives the same information but in averages of only 
27 pulses.  The breadth of the disorderly Q-mode emission is very clear 
in this figure as is the brightness of the occasional intense Q-mode pulses 
which contribute to the averages that are plotted.  The two bright features 
between pulses 5500 and 6000 are probably caused by interference.  The 
lower panels in each figure give the profile computed over the full length 
of the observation and are thus identical.  Note that this profile is double 
attesting to the strength of the transient second component for a substantial 
period after the modal transition.  

Remarkably, the two other Q- to B-mode transitions identified in our 
observations behave very similarly.  The one on 12 July occurred near 
pulse 5266 rather late in that day's observation, and a higher level of 
interference on 4 October initially complicated our finding the event there 
at pulse 1755.  The left-hand portion of Figure~\ref{fig5} shows how the 
profile changes in all three transitions track each other.  Here we plot the 
ratio of the component-height maxima A(2/1) as a function of profile number, 
where each profile consists of 256 successive pulses.  The three transitions 
are aligned such that profile 1 of each begins with its first B-mode pulse.  
Note how closely the three first profiles resemble each other, all with their 
second component peak amplitudes a remarkable 1.75 times their first!  The 
diagram shows considerable scatter within the first 15 or so profile amplitude 
ratios, but overall one has the impression of an exponential-like behavior.  
And it is useful to keep in mind that profile 15 follows the B-mode onset by 
some 70 minutes!

\section*{V. Variation of the `Carousel' Circulation Time $\hat{P_3}$}

\begin{figure}
\begin{center}
\epsfig{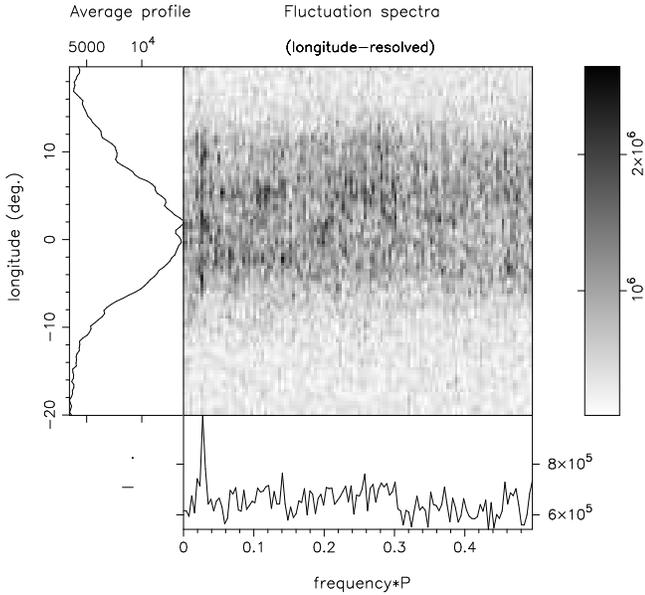}
\end{center}
%\parhang {\bf Figure 6.}---
\caption{Longitude-resolved fluctuation spectra for the initial Q-mode interval of the 2003 March 10 observation.  The prominent feature in the bottom panel at 0.0275$\pm$0.001 c/$P_1$ corresponds to an interval of 36.4$\pm$0.9 $P_1$---apparently the first detection of a Q-mode circulation time. Note also the complete lack of a B-mode ``drift'' feature at about 0.45 c/$P_1$ and the broad, single Q-mode profile centered at zero longitude.  These MJD 52709 spectra were computed using the first 2048 pulses, whereas the B-mode recommences at about pulse 2540.
\label{fig6}}
\end{figure}

In Paper III we noted interesting correlations between the forms of the B-mode 
profiles and the circulation times computed from fluctuation spectra of their 
constituent PSs.  The various observations in that paper, though extensive, 
were of inadequate length to see how these variations came about.  Moreover, 
it was unclear whether subbeam circulation persisted in the Q mode; we found 
some hints in Paper I that it did do, but no Q-mode PS then in our possession 
exhibited a fluctuation feature demonstrating such circulation.  

We were thus surprised and gratified when the Q-mode PS preceding the B-mode 
onset of 2003 March 10 exhibited a strong low frequency modulation feature.  The 
longitude-resolved fluctuation (hereafter lrf) spectra are given in Figure~\ref{fig6}.  
They are computed from the first 2048 pulses in the PS having a B-mode onset at 
pulse 2540.  A ``boxcar'' smoothing of 3 samples has been applied to the PS to 
enhance the S/N of the spectra.  The Q-mode profile is clearly recognizable in the 
left-hand panel and the integral spectrum is shown in the lower panel where a 
strong feature at some 0.025 cycles/period (hereafter c/$P_1$) is evident.  Careful 
measurement of the frequency of this feature results in a value of 0.0275$\pm$0.001
c/$P_1$.  This is the first time that a fluctuation feature corresponding to the 
circulation time has been seen in our meter-wavelength observations, though such 
a feature has at times been seen in the decameter PSs of Paper II.  The circulation 
time associated with this fluctuation frequency is then some 36.4$\pm$0.9 
$P_1$---a somewhat lower value that usually found for B-mode circulation times.  

\begin{figure}
\begin{center}
\epsfig{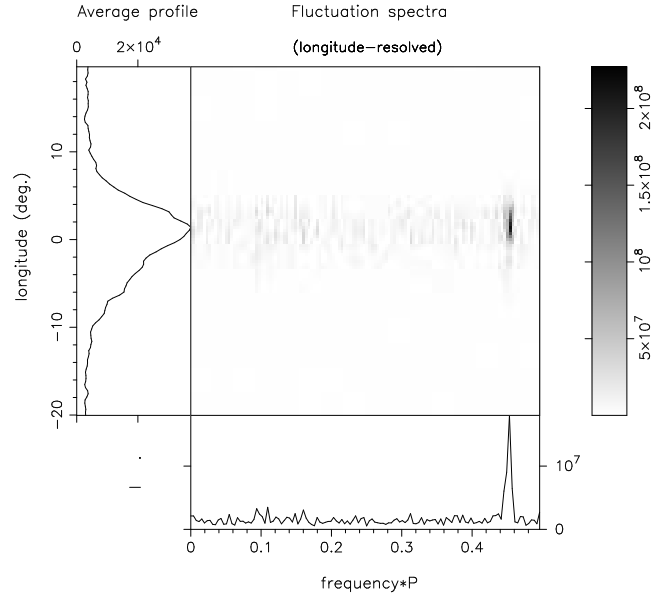}
\end{center}
%\parhang {\bf Figure 7.}---
\caption{Longitude-resolved fluctuation spectra computed from 256 pulses just 
after the B-mode onset of the 2003 March 10 observation. Here the ``drift'' feature 
frequency is 0.452 c/$P_1$, which corresponds to a 20-subbeam ``carousel'' 
circulation time of 36.47$\pm$0.03 $P_1$---just the same as that determined for 
the Q-mode in Fig.~\ref{fig6}. 
\label{fig7}}
\end{figure}

Subbeam circulation times can more readily be computed for B-mode PSs from 
the strong roughly 0.46 c/$P_1$ subpulse-modulation feature always associated 
with this behavior.  We further know that this response is a first-order alias of the 
true fluctuation frequency of about 0.54 c/$P_1$---a matter which has been 
demonstrated extensively using the harmonic-resolved fluctuation spectrum (\eg, 
Paper I).  Finally, every analysis of the star's B-mode properties has resulted in the 
conclusion that the observed subpulse drift is produced by a circulating pattern 
of just 20 subbeams.  Thus, the B-mode circulation time can be computed as 
$\hat{P_3}$=20$P_3$(true)=$20/[1-f_3({\rm obs})]$.  

We give an example of such a computation in Figure~\ref{fig7} for the 256 pulses 
immediately after the B-mode onset of the 2003 March 10 observation at pulse 2540.  
Here the highly unusual profile just following the B-mode onset is given in the 
left-hand panel, and the bottom panel shows the integral spectrum associated 
with the drifting-subpulse modulation.  The feature frequency is here some 
0.452$\pm$0.003 c/$P_1$---one of the smallest such values ever measured for the 
star's B mode.  

\begin{figure}
\begin{center}
\epsfig{file=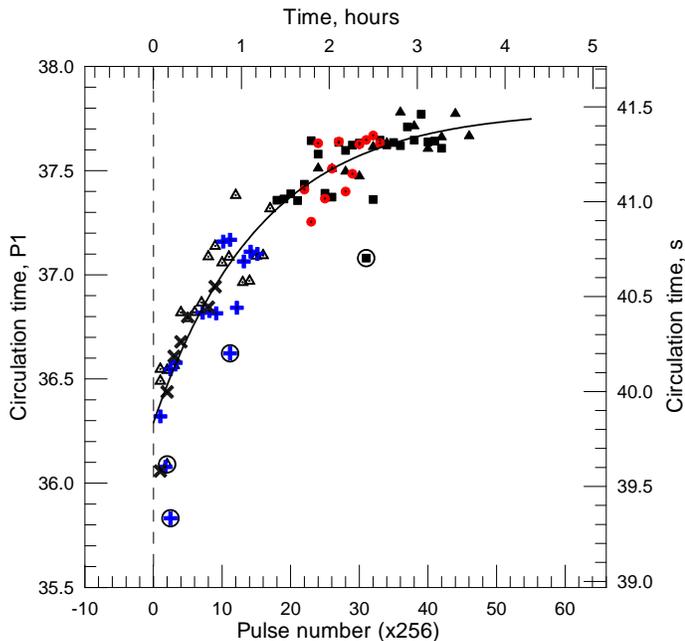,height=8.5cm,angle=0}
\end{center}
%\parhang {\bf Figure 8.}---
\caption{Subbeam ``carousel'' circulation time $\hat{P_3}$ as a function of 
time from the B-mode onset at intervals of 256 $P_1$; 1 hour is equal to 12.8 
such units. Circulation times for the three Q- to B-mode transition observations 
are aligned at the origin (vertical dotted line) and fitted with the exponential 
curve (solid line). The other three ``pure B''-mode observations could be 
fitted to this curve by shifting along the  time axis at 18, 24 and 22 (256-$P_1$ 
units), respectively. The characteristic time is some 4000 pulses or 1.22 
hours. It is also about 107 circulation times. All three days with ``pure 
B''-profiles show a gradual $\hat{P_3}$ increase from some 37.3 to about 
37.8 $P_1$ during the observational session and thus seem to be a later part 
of the Q- to-B transformation process. Four of the lrf spectra exhibited double 
peaks and $\hat{P_3}$s calculated for these weaker precursor peaks are 
separately indicated (open circles) but not included in the fitting procedure 
(see text). 
\label{fig8}}
\end{figure}

Generally, it was possible to measure a circulation time for each 256-pulse interval 
following the respective B-mode onsets in each of the three observations.  Not only 
do the $\hat{P_3}$ values of these three transitions behave similarly as a function 
of time, but they too suggest an exponential dependence of the form $[1-\exp(-t/\tau)]$.  
The behavior of all three observations is plotted in the left-hand part of Figure~\ref{fig8}, 
where the three observations are depicted with triangle, + and $\times$ symbols.  
Note as well that several other of the initial circulation-time values fall at just more 
than 36 $P_1$, thus agreeing well with the more poorly determined Q-mode value 
from Fig.~\ref{fig6}.  

Remarkably, we can then apparently conclude that the initial B-mode, subpulse-drift 
associated circulation time represents a continuation of that associated with the very 
disorderly emission of the star's Q mode.  

%Comment: You can choose between fig8 and 8a. 
%In Figure~\ref{fig8} Y axis is given in $P_1$, fitting ln(Y)=0.0112*ln(X)+3.566; n=88, Coef of determination, R-squared=0.926. The four open circles in Fig.~\ref{fig8} represents ``accelerated'' subgroups inside basic groups of 256 pulses which manifest themselves as secondary peak in fluctuation spectra. The average CT for pure B-modes are 37.52$P_1$ (B-3), 37.55$P_1$ (B-1) and 37.62$P_1$ (B-2). 
%In fig 8a Y- axis is given in seconds .
%Fitting line Y(n)=0.0112*ln(X)+3.65. n=88.
%  It is intriguing that for all 3 pure B-modes CT increases with time. Energy ratio of two subwindows W1/W2  as welll as peak intensity (fig10) increases with time too. It is likely that, especially evidently for B1-day (52711).
%Comment: Linear fit: Y=-0.70*X+26.74; n=86. Coef of determination, R-squared=0.72
%High degree of correlation allow making estimation of CT \& drift rate just using pulse components ratio. 

\section*{VI. Locating the ``Pure B''-mode Observations Within the Transition Recovery}

We then turned our attention to the analysis of the three long PSs in which only 
B-mode emission was observed, those recorded on 12 March, 20 July and 5 
October, respectively.  Each PS exhibited two properties which surprised us 
initially:  a) there was a perceptible tendency for the amplitude ratio A(2/1) 
to decrease, and b) we also found strong evidence in two of the three for a 
lengthening of the circulation time over their duration.  Of course, we now 
understand that a Q- to B-mode transition had to occur somewhat before the 
beginning of our data-acquisition in each of these cases---the only question is 
just how long before.  

The profile and circulation-time variations give us an indication of just where the 
``pure B''-mode observations fall in relation to the B-mode onset.  If we proceed 
on the strong appearance that these quantities do exhibit a $[1-\exp(-t/\tau)]$ 
functional dependence, then we can locate the various average profiles 
and their corresponding $\hat{P_3}$ values appropriately and then partially test 
whether the procedure was correct.  Indeed, we find again that the two measures 
vary together such that the three days can be interpolated at a time corresponding 
to the 18th, 24th and 22nd 256-pulse average after the B-mode onset, respectively.  
We have plotted the A(2/1) values in Fig.~\ref{fig5} and the $\hat{P_3}$ 
measurements in Fig.~\ref{fig8} as is also described in their captions.  

While the above interpolation procedure must be regarded as approximate, it 
also cannot be wildly wrong.  The above circulation-time values following the 
B-mode onset were fitted to a $[1-\exp(-t/\tau)]$ function, which is shown by the 
continuous curve in Fig.~\ref{fig8}.  The characteristic time $\tau$ is then about 
73 minutes or 4000 $P_1$.  It is also some 107 circulation times.  In terms of the
light-time scale of the pulsar's magnetosphere, to say nothing of the acceleration 
regions within its polar flux tube, this is a very long time indeed.  It is hard to 
imagine what physical processes associated with pulsar emission could have 
such a long time scale.  

\begin{figure}
\begin{center}
\epsfig{file=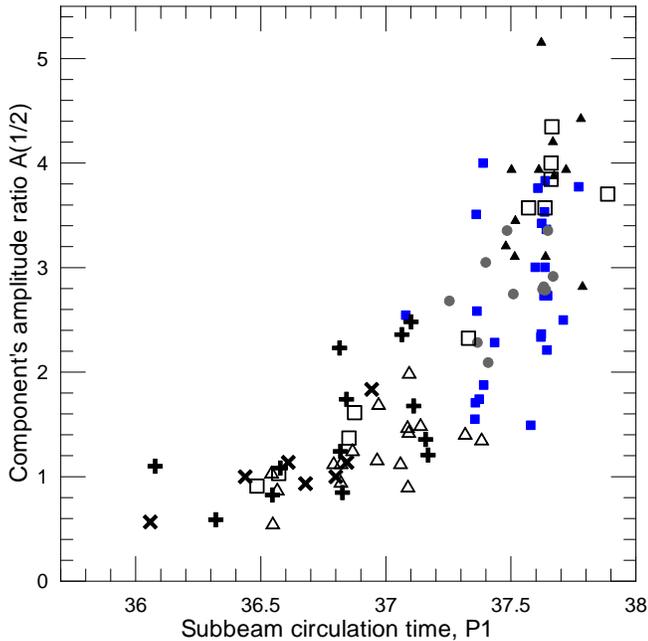,height=8.5cm,angle=0}
\end{center}
%\parhang {\bf Figure 9.}---
\caption{Component amplitude ratio A(1/2) as a function of subbeam 
circulation time $\hat{P_3}$ at 327 MHz. The 103-MHz values (open 
square symbols) were taken from Table 1 of Paper III.
\label{fig9}}
\end{figure}

Finally, in Figure~\ref{fig9} we plot the component peak-amplitude ratio A(1/2) 
versus the subbeam-circulation time $\hat{P_3}$ for all of six observations as 
well as some further observations from Paper III (open square symbols).  The 
values are well fitted by an inverse relation.  Overall, it is clear that the 
subbeam-circulation time and profile form are closely related, such that one 
could reliably be estimated from the other.

\section*{VII. Asymmetry in the Mode-Changing Process}

We see good evidence from the foregoing analyses that the modal-transformation 
process of the emitting regime in pulsar B0943+10 is not symmetrical in time.  
Thus far we have only managed to observe one B-to-Q-modal transition---this 
the remarkable 18-min (986-$P_1$) Arecibo 430-MHz observation obtained 
in 1992, which was thoroughly investigated in both SIRR and Paper I.  

Comparison of the evolution of the pulse shape and subbeam-carousel rotation 
time $\hat{P_3}$ for the respective Q-to-B (2003) and B-to-Q (1992) transitions 
reveals significant differences---\ie, in the latter case these processes occur 
much more quickly.

\begin{figure}
\begin{center}
\epsfig{file=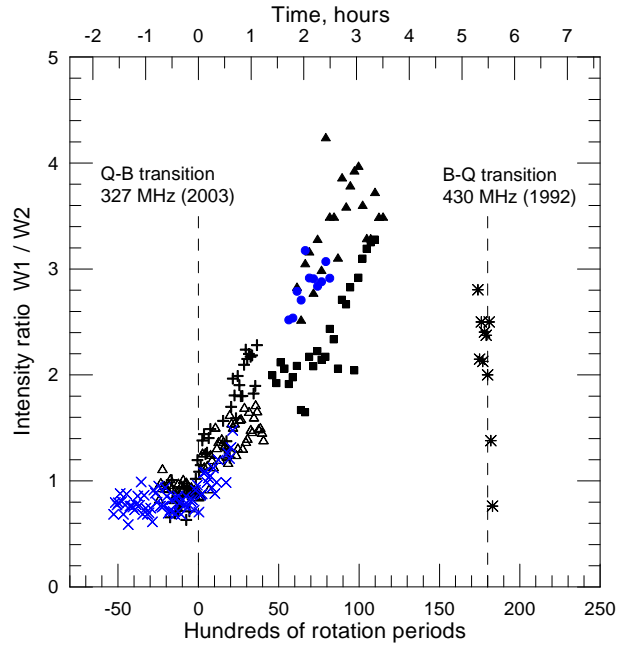,height=8.5cm,angle=0}
\end{center}
%\parhang {\bf Figure 10.}---
\caption{
Intensity ratio of two emission sub-windows for negative longitudes W1 
and positive longitudes W2 as a function of time following the B-mode 
onset in units of the group number of successive 100-pulse averages. 
The pulse shape is apparently stable during the 1.5 hours of Q-mode 
emission, whereas it continuously changes after the B-mode 
recommencement. The mode transition times are shown by vertical 
dashed lines. The shift of the 1992 B- to Q-mode observation (430 MHz) 
along time axis is arbitrary. This figure clearly shows the different pulse-shape 
transformation rate for Q-to-B- (slow) and B-to-Q-mode (fast) transitions 
in B0943+10. 
\label{fig10}}
\end{figure}

To investigate the pulse-shape evolution we divided the pulse window onto 
two subwindows W1 and W2 at negative (--8\degr, 0) and positive longitudes 
(0, +8\degr). As was shown earlier at 430 MHz, the maxima of the B-mode and 
Q-mode profiles are separated in longitude at a distance nearly equal to the  
half-power width of the B-mode profile. This means that a longitude on the 
trailing edge of the B-mode profile corresponding to the half-power point can be 
taken as a zero longitude, equally remote from regions where the dominance 
of each mode is strongest. Temporal variations of the intensity ratio in these 
two subwindows W1/W2 are presented in Figure~\ref{fig10}.  The intensity of 
the subwindow for each individual pulse is obtained by summing the amplitudes 
of all the bins for a given subwindow---besides additionally averaging over 100 
adjacent pulses to diminish the influence of the noise and possible scintillation 
fluctuations. Note that the W1/W2 ratio changes from 2.8 to 0.8 during the 18-min 
B-to-Q profile-shape transformation; whereas in the opposite Q-to-B transition 
the same change in pulse shape takes about 2.5 hours!   Similarly, the $\hat{P_3}$ 
value at the B-mode recommencement is just over 36 $P_1$, while at the Q-mode 
onset it was about 37.35 $P_1$.  One can suggest that a sharp decrease of the 
circulation time down to 36 $P_1$ occurs after Q-mode commencement, without 
further significant changes of the pulse shape. Note that the W1/W2 ratio is nearly 
constant (W1/W2 is randomly distributed in the range 0.6-1.0) during the initial 
46-min (2540 pulses) Q-mode interval of 10 March and the 96-min (5266 pulses) 
Q-mode PS of 12 July. 

%Comment to fig.10: On fig 10 the shift of the B-days in respect of B-mode recommencement at pulse \#2540 (MJD 52709) is the same as in fig 8 or 8a.
%25.40+(30*256/100)=102.2
%25.40+(43*256/100) =135.48,
%The B-mode recommences at about pulse 25.40 (dashed vertical line on fig 10).
%The 430-MHz data are shifted arbitrary.

\section*{VIII.  Discussion}

Pulsar B0943+10's circulation time $\hat{P_3}$ in Fig.~\ref{fig8} gradually 
changes in the range from about 36 $P_1$ to at least 37.8 $P_1$ following 
the onset  of its B mode.  The difference (1.8/36=0.05) may be as large as 
5\%. 

Drift-rate variations are well known to occur in a number of pulsars, and 
we now see that B0943+10 is another member of this group.  Some stars 
such as B0031--07 (\eg, Wright \& Fowler 1981a; Vivekanand \& Joshi 1997), 
B1944+17 (\eg, Deich \etal\ 1986), B2303+30 (Redman \etal\ 2005) and 
B2319+60 (\eg, Wright \& Fowler 1981b) exhibit several discrete drift-rate 
``states'' that are usually designated as modes.  Others show what appear 
to be continuous drift-rate changes from rarely (\ie, B0809+74; van Leeuwen 
\etal\ 2002) to very frequently (B0826--34 and B2016+28)---and in B0031--07
and B2303+30 a regular cycle has been detected following occasional 
``events'' wherein ``null'' pulses are often observed to accompany drift-rate 
variations.  

While the discrete drift rates can change by several factors of two, the 
amplitude of observed continuous drift-rate variations is less than 8\% as 
in B0943+10, and their characteristic times some several minutes---\ie, 
significantly less than we observe in B0943+10.  Moreover, B0943+10 
surely provides the best example of drift-rate changes which correlate with 
large changes in the profile form. In most other such pulsars drift-rate 
variations are associated only with smaller and less orderly changes in 
the amplitude, shape and/or phase of the pulse profile. Systematic variations 
of the drift rate in B0826--34 (and probably B2303+30) apparently cross 
alias boundaries and thus produce changes in the drift direction (Esamdin 
\etal\ 2004; Gupta, Gil, Kijak, \& Sendyk 2004). An even stronger similarity 
with B0943+10 is the discovery by Esamdin \etal\ that B0826--34's long 
intervals of ``null'' pulses are in fact a weak Q-like mode, which has a very 
different profile from that of the strong, drifting mode (see their fig. 2). This 
weak mode lasts longer than the stronger mode, and its profile is also found 
to be stable in intensity and shape.  Moreover, both stars are now known 
to have closely aligned geometries.

In B0943+10 it is notable that null pulses have not been detected despite 
its prominent drift-rate variations.  In its bright B mode an upper limit on 
the number of nulls might be placed at around 0.1\%, but in the chaotic 
Q mode the intensity range of individual pulses is so great that no 
sensible discrimination can yet be made between possible nulls and 
weak pulses.  
 
An effort to understand drift-rate variations in the context of revisions to 
the polar-cap accelerator theory of Ruderman \& Sutherland (1975) has 
recently been published by Gil, Melichidze \& Geppert (2003).  These 
authors attribute such variations to the shielding effect of iron ions whose 
flow is sensitive to the surface temperature of the polar cap, which in 
turn is heated by the bombardment of ultra-relativistic charged particles
originating from pair production. The model implies that the cooling 
process of the star's surface below the mean temperature leads to a 
faster than mean drift rate, while the reverse process of heating results  
in slower than mean drift rates. The drift-rate and direction changes of
B0826--34 at 325 MHz were discussed by Gupta, Gil \etal\ (2004). A 
typical time scale for one cycle of drift-rate reversal is about 100 stellar 
rotation periods, or nearly 3 min for this pulsar.  In B0826--34 drift-rate 
increases (corresponding to cooling below the mean temperature) are 
found to occur more slowly and smoothly than drift-rate decreases. 

In B0943+10 the cycle is just the opposite: the slowly decreasing 
B-mode drift rate (corresponding ostensibly to heating above the mean 
temperature) is the slower process. Thus, this mechanism would appear 
to be backward for B0943+10, contrary to B0826--34.  Nevertheless, it is 
clear that the circulation-time increases and decreases of the subbeam
carousel have different characteristic times for the two pulsars.  The 
current theory predicts very short cooling time scales (a few $\mu$sec 
for magnetic fields strengths of a few 10$^{12}$ G; Gil \etal\ 2003) that 
could explain microstructure of individual pulses or short nulls.  That 
any physical mechanism could cause multi-hour-scale variations in the 
polar cap temperature in pulsars is still very much open to question.

As mentioned above, a number of pulsars exhibit drift-rate variations 
in a range up to about 8\% like B0943+10. None of the others, however, 
demonstrate B0943+10's pronounced and correlated changes in profile 
form.  Though we cannot be sure what agent or mechanism is causing 
these profile variations, we are attracted by the analogy to the average 
pulse-shape changes in two well studied ``mode switchers'', B0329+54 
and B1237+25.  Core emission is prominent in these two stars, surely 
in substantial part because our sightline passes much closer to their 
respective magnetic axes, but in both cases clear modal variations are 
also seen in the characteristics of their conal components.  A tendency 
for profile modes to represent more and less symmetrical configurations 
of emission (about the longitude of the magnetic axis) has long been 
noted (\eg, Rankin 1986), and the contrasting properties of the modes 
in these two stars is particularly striking in this respect:  both exhibit 
roughly symmetrical profiles in their respective ``normal'' modes and 
then strong leading/trailing emission asymmetries accompanied by 
enhanced core activity in their ``abnormal'' modes.  In B0329+54 at 
meter wavelengths (\eg, some 111--606 MHz), mode changes are 
marked by alterations in the positions and intensities of the core and 
leading conal components (Gangadhara \& Gupta 2001; Suleymanova
\& Pugachov 2002), and a very similar pattern is observed in B1237+25 
(Bartel \etal\ 1982; Srostlik \& Rankin 2005) as well apparently as in   
B0355+54 in the course of its slow modal variations (Morris \etal\ 1980).

In B0943+10 [as in other conal single (${\bf S}_d$) stars] our sightline 
cuts across the periphery of its conal emission pattern, so we can have 
no direct indicator of its core radiation intensity.  Nevertheless, we can 
hypothesize that B0943+10 belongs to the group of pulsars with variable 
core activity.  In B0329+54 and B1237+25, core-active intervals are 
observed as sudden eruptions, whereas in B0943+10, this activity may 
increase gradually over several hours.  As a result the star's symmetrical 
profile (seen in its Q mode and also just after B-mode onset) evolves to 
its usual, highly asymmetrical form several hours into each B-mode interval.  
It is also possible that the profile change at B-mode recommencement 
from ``single'' to ``double resolved'' entails a conal emission-height 
increase that is prompted by core activity (indeed, a possibility worth 
exploring for B0826--34 as well).  

Recent XMM-Newton observations by Zhang \etal\ (2005) have detected 
X-ray emission from B0943+10, and several current models suggest 
that such radiation is basically thermal, emitted from a polar cap surface   
heated by back-flowing plasma produced by the same pair-processes  
that are responsible for the star's radio emission.  These authors conclude 
that ``within the thermal interpretation, the X-ray radiation is emitted from 
a heated area much smaller than the conventional polar cap area''.  This 
result, however, was based on a 20-ks observation---one surely long 
enough to involve one or another (or a mixture) of B and Q-mode intervals.  
Increased core activity (associated with one of the modes) might alter the 
polar-cap surface temperature and distribution (if the cooling time-scales 
are indeed long enough), affecting ${\bf E}\times$${\bf B}$ and/or the conal 
beam symmetry as mentioned above.  The above result takes no account 
of this and is thus of uncertain significance.  

If, however, the ``feedback'' model of Wright (2003) is substantially 
correct, the situation may be much more complex.  Here the ``carousel''
circulation time would depend on the ``connectivity'' between the polar 
cap and ``outer gap'' accelerators, and only indirectly on the 
${\bf E}\times$${\bf B}$ drift.  In this aurora-like scenario, the variation in 
both rotation speed and profile asymmetry would be produced by the 
inevitably asymmetric interaction between the magnetic poles of an 
inclined rotating star. Even in a perfect dipolar geometry, the slowly 
circulating emission columns at either pole would not remain on 
matching fieldlines and their communication paths would gradually 
become ``twisted'' in a manner not unlike the earlier suggestion of Paper III.

\section*{IX. Summary and conclusions}

B0943+10's subpulse-drift rate is observed to decrease exponentially by 
some 5\% during the 4+ hours after B-mode onset with a characteristic 
time of some 73 min---and a very similar behavior is seen in all three 
327-MHz observations.   These drift-rate changes are accompanied by 
changes in the integrated pulse shape. Remarkably, the trailing component 
brightens at B-mode onset, making the overall profile a conal double, and   
for the first 20 mins or so at 327 MHz it outshines the usual leading emission 
feature.  Overall, the amplitude ratio of the two B-mode profile components 
varies significantly from 1.75 to 0.2 during the same 4+ hours. Apparently 
this slowly decreasing drift rate eventually falls below some critical value, 
and is replaced by the disorderly drift pattern of the alternative Q emission 
mode.  This Q mode is characterized by a broad Gaussian-shaped profile 
that is stable over several hours, a complete absence of regular subpulse 
drift, and a diminuation to about half in emission intensity.  This highly 
organized process of gradually decreasing subpulse drift rate accompanied 
by systematic profile variations has no simple parallel in other radio pulsars.  
In particular, no other known pulsar exhibits systematic variations on a time 
scale even approaching the 1.2 hours seen in B0943+10.  

Our main results can be summarized as follows:
\begin{itemize}
\item Six observations at 327 MHz with a total duration of more than 11 hours 
were carried out in 2003 in order to search for modal transitions.
\item Three of the days exhibited Q- to B-mode transitions and three others 
were found to have long B-mode sequences.
\item All three Q- to B-mode transitions appear to behave very similarly, 
qualitatively and quantitatively.
\item A Q-mode circulation-time feature was detected for the first time, and 
its value is some 36.4$\pm$0.9 $P_1$.
\item The B-mode circulation time is found to be some 36.2 $P_1$ just after 
its onset.
\item The B-mode circulation time exhibits an exponential relaxation to an 
asymptotic value of perhaps 37.8 $P_1$ with a characteristic time of 1.2 
hours.
\item The early B-mode profile is found to be double---with a somewhat 
larger second component---which then gradually decreases over a several
hour period, such that the magnetic-axis longitude falls at the trailing 
half-power point of the asymptotic B-mode profile. 
\item The symmetrical to asymmetrical subbeam-carousel visibility---demonstrated 
by the dramatic B-mode profile variations---points to a complex physical 
origin related to the ``absorption'' phenomenon.
\item Q-to-B-mode changes in pulsar B0943+10 again have a rapid and 
very slow aspect.  The 1.2-hour characteristic time may be the longest 
orderly behavior known to be associated with pulsar emission.
\end{itemize}

\bigskip

\noindent {\bf Acknowledgements:} We thank Geoff Wright for his 
critical contributions to the paper in draft.  This work was also supported 
in part by US NSF grants AST 99-86754, 00-98685 as well as (for JMR) 
a visitor grant from the Nederlandse Organisatie voor Wetenschappelijk 
Onderzoek.  Arecibo Observatory is operated by Cornell University under 
a cooperative agreement with the US NSF.

\end{document}